\documentclass[a4paper,UKenglish,cleveref, autoref, thm-restate]{lipics-v2021}
\pdfoutput=1 
\hideLIPIcs  

\usepackage{multirow}
\graphicspath{{./images/}}
\usepackage{import}
\title{Type-safe and portable support for packed data}

\author{Arthur Jamet}{University of Kent, Canterbury, UK}{aj530@kent.ac.uk}{https://orcid.org/0009-0005-5230-0667}{}

\author{Michael Vollmer}{University of Kent, Canterbury, UK}{m.vollmer@kent.ac.uk}{https://orcid.org/0000-0002-0496-8268}{}

\authorrunning{A. Jamet and M. Vollmer} 

\Copyright{Arthur Jamet and Michael Vollmer} 

\ccsdesc[300]{Information systems~Data layout}
\category{Experience Paper}

\keywords{program optimisation, data structures, data layout, packed data} 

\relatedversion{} 
\nolinenumbers

\begin{document}

\maketitle
\newcommand{\packedhaskell}[0]{\lstinline@packed-data@ }
\newcommand{\var}[1]{\lstinline|#1|}
\newcommand*\circled[1]{\raisebox{.5pt}{\textcircled{\raisebox{-.9pt} {#1}}}}
\begin{abstract}
	When components of a system exchange data, they need to serialise the data so that it can be sent over the network. Then, the recipient has to deserialise the data in order to be able to process it. These steps take time and have an impact on the overall system's performance.

A solution to this is to use \textbf{packed data}, which has a unified representation between the memory and the network, removing the need for any marshalling steps. Additionally, using this data representation can improve the program's performance thanks to the data locality enabled by the compact representation of the data in memory.
Unfortunately, no mainstream programming languages support packed data, whether it's out-of-the-box or through a compiler extension.

We present \packedhaskell, a Haskell library that allows for type safe building and reading of packed data in a functional style. The library does not rely on compiler modifications, making it portable, and leverages meta-programming to allow programmers to pack their own data types effortlessly.

We evaluate the usability and performance of the library, and conclude that it allows traversing packed data up to 60\% faster than \textit{unpacked} data. We also reflect on how to enhance the performance of library-based support for packed data.

Our implementation approach is general and can easily be used with any programming languages that support higher-kinded types.

\end{abstract}
\section{Introduction}

Components within a system may need to exchange data. Consequently, data needs to be sent in a format that is processable by the recipient. For example, web application programming interfaces (web APIs) usually serialise data using the Java-Script Object Notation (JSON) before sending it over the network.
On the other hand, the receiver of the data has to \text{deserialise} the response into a processable object; only then will it be able to use the data. For example, a web client has to parse the JSON received before being able to, say, compute the average length of the usernames in a list.

The most common serialisation formats used for these kinds of interactions are JSON, XML and HTML. They all have a simple, human-readable syntax, and it is very easy to serialise and deserialise data with them. However, because of their syntax, these formats make the data considerably larger, leading to a higher memory footprint and longer transfer times. Additionally, de/serialisation has a cost and can impact the performance of large exchange-heavy systems.

Alternatively, such components could use a binary format to exchange data. For example, one could use Google's Protocol Buffers (Protobuf)\footnotemark{}\footnotetext{https://protobuf.dev}, which generates code (for many languages) to serialise/deserialise data to/from a binary format. This leads to smaller payloads (compared to human-readable formats), and thus smaller transfer times. On the other hand, Cap'n Proto\footnotemark{}\footnotetext{https://capnproto.org} allows using binary data as-is, without any marshalling steps, but its encoding leads to bigger buffers compared to Protobuf's.

\textit{Packed} data is binary data that can be used as-is, safely and without any deserialisation step. Thus, it can be easily persisted to and read from the disk without any conversion cost. Additionally, the minimal layout information contained in the packed data leads to a smaller overhead compared to Cap'n Proto.
Moreover, unlike objects in memory, packed data does not have pointers and nested fields are inlined within the parent structure. Consequently, the data locality enabled by the packed format can leverage better performance (e.g.\ using the L1 cache). This can help speed up programs that traverse large data trees, like compilers with abstract syntax trees (AST) or web browsers with document object models (DOM).

\begin{figure}[h]
	\begin{subfigure}[b]{0.5\textwidth}
		\centering
		\captionsetup{justification=centering}
\begin{lstlisting}[language=C]
enum Tag { LEAF, NODE };

struct Tree {
  enum Tag tag;
  union {
    // Content of a leaf
    struct {
      long value;
    };

    // Content of a node
    struct {
      struct Tree *left;
      struct Tree *right;
    };
  };
};
\end{lstlisting}
		\caption{Definition of the tree, in C}
		\label{fig:c-definition}
	\end{subfigure}
	\begin{subfigure}[b]{0.49\textwidth}
		\centering
		\captionsetup{justification=centering}
		\includegraphics[width=\textwidth]{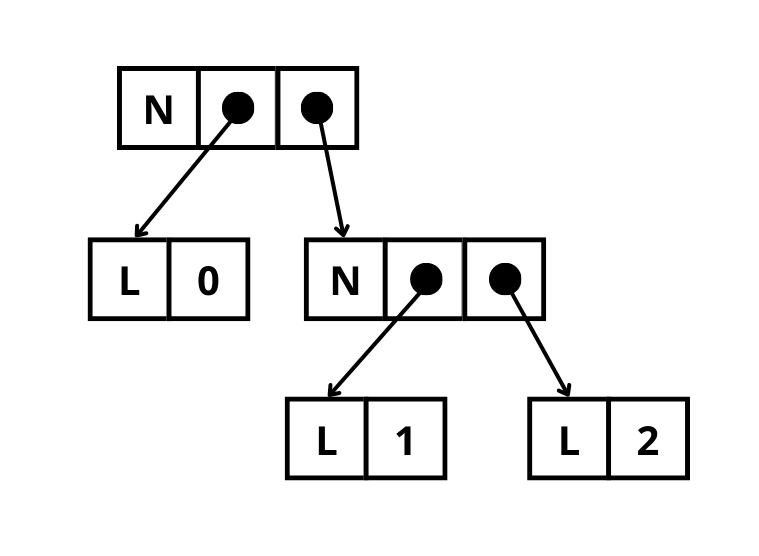}
		\caption{Pointer-based representation}
		\label{fig:unpacked-layout}
	\end{subfigure}
	\begin{subfigure}[b]{0.5\textwidth}
		\centering
		\captionsetup{justification=centering}
		\begin{lstlisting}[language=Haskell]
data Tree = Leaf Int 
          | Node Tree Tree
		\end{lstlisting}
		\caption{Definition of the tree in Gibbon/Haskell}
		\label{fig:packed-definition}
	\end{subfigure}
	\begin{subfigure}[b]{0.49\textwidth}
		\centering
		\captionsetup{justification=centering}
		\includegraphics[width=0.8\textwidth]{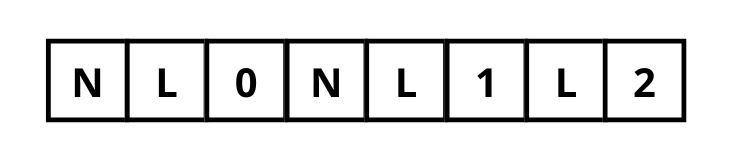}
		\caption{Packed representation}
		\label{fig:packed-layout}
	\end{subfigure}
	\caption{Pointer-based and packed representation of a data tree}
	\label{fig:packed-vs-unpacked}
\end{figure}

The current state-of-the art is Gibbon, a compiler that transforms programs that use pointer-based structures into programs that use this packed representation~\cite{gibbon,local,parallel-gibbon,marmoset}. Gibbon supports a small subset of the Haskell programming language, and compiles recursive functional programs operating on trees into C code operating on packed bytestrings. For example, Figure~\ref{fig:unpacked-layout} shows the memory layout of a binary tree (defined in Figure~\ref{fig:c-definition}) for a C program. Gibbon's packed representation of that same tree (defined in Figure~\ref{fig:packed-definition}) is given in Figure~\ref{fig:packed-layout}: instead of containing pointers, the parent node contains its children directly, contiguously in memory.

Writing C code for traversing packed data by hand is not trivial, and requires raw and untyped pointer manipulations, leading to code that is hard to read and maintain. Figure~\ref{fig:c-packed-sum} gives an example of such code, which computes the sum of the leaves in a packed tree. Using a specialised compiler like Gibbon can alleviate this concern by generating the tricky code automatically (and using a typed intermediate language that verifies packed data layout invariants~\cite{local}), but introducing a dependency on an experimental research compiler is not realistic for some software projects. Additionally, Gibbon supports only a small subset of Haskell, lacking many convenient features, and also requires whole-program compilation, making integration into larger programs awkward.

\begin{figure}[ht]
\begin{lstlisting}[language=C]
long sum(char **t) {
  long res = 0;
  if (**t == LEAF) {
    *t = *t + 1;
    res = *((long *)*t);
    *t = *t + sizeof(long);
  } else if (**t == NODE) {
    *t = *t + 1;
    res = sum(t);
    res = res + sum(t);
  }
  return res;
}
\end{lstlisting}
	\caption{Computing the sum of the values in a packed tree, in C}
	\label{fig:c-packed-sum}
\end{figure}

This data layout has another a drawback: it does not allow accessing members of a data structure easily. Since fields are inlined and can have variable sizes (e.g. for recursive data types), they do not have a predefined position (or offset) within the parent structure.
For example, consider a tree similar to the one from Figure~\ref{fig:packed-vs-unpacked}, but where the root node's children are swapped. We know that the leaf is \textit{after} the node, but we do not know the node's size or where it ends. The only way to access the leaf would be to traverse the first child. Depending on the size of that tree, it can lead to catastrophic performance issues.
One solution to this problem is to use \textit{indirections}. In Gibbon, indirections are numbers inserted (when requested) in the packed buffer before each field of a data structure. These numbers represent the size in bytes of the field they precede~\cite{local}.
Indirections make the packed data larger, but they help avoid useless traversals of (potentially big) preceding fields before getting to the target one.

While the benefits of packed data have been proven, Gibbon is one of the only compilers that use this data layout natively.
But Gibbon is not the only tool to leverage a condensed data layout: we already mentioned high-level libraries like Cap'n Proto, which allows using packed data as-is, and Protobuf, to pack/unpack data. On the compiler-level, the Glasgow Haskell Compiler (GHC) also supports packing data, through \textit{compact normal forms} (CNF)~\cite{cnf}, but we can't manipulate it (e.g. access a field) in this form.
Packed data remains to be leveraged by mainstream programming languages. The performance improvements enabled by this data format remain limited to experimental research projects.

In this paper, we present a new effort of supporting packed data through a library. \packedhaskell is a Haskell library for building and traversing packed data. Inspired by the example packed data interface from Bernardy et al.'s paper~\cite{linear-haskell}, the library contains a code generator that allows the programmer to easily use their own algebraic data types (ADT). We use Haskell in this paper, but our approach is not specific to the Haskell programming language. Thus, the library does not require any compiler modifications and can be implemented in any language with higher-kinded types.
We present the design of this library along with a use-case for packed data (Section~\ref{sec:packed-haskell}). Then, in Section~\ref{sec:eval}, we evaluate the usability and performance of our library-based approach. Finally, we reflect on the limitations of our design in Section~\ref{sec:reflection} and suggest improvements in Section~\ref{sec:future-work}.

The source code of \packedhaskell is available on Hackage\footnote{https://hackage.haskell.org/package/packed-data}. The package also includes the code of the benchmarks we present in Section~\ref{sec:bench}.

Unless specified, all code snippets in this paper use the Haskell syntax.

\section{The \packedhaskell Library}
\label{sec:packed-haskell}

In this section, we will go through and explain the features of the \packedhaskell library. Then, we will give an example of a writing operation on packed data, and finally we will cover a use-case of packed data, implemented with the library.

\subsection{Features}

As its name suggests, \packedhaskell allows creating and manipulating packed data. Since the library is implemented in Haskell, we leverage the language's type-system to guarantee that packed data manipulation is type-safe. Thus, any bad use of a packed buffer will be detected and rejected at compile-time.
Additionally, we hide pointer arithmetic details behind a functional API. This allows programmers to use packed data like they would use data streams, instead of a raw and untyped memory buffer.
The library also includes a code generator which generates all the necessary functions to pack, traverse and unserialize custom ADTs.
Finally, \packedhaskell allows programmers to choose whether indirections are inserted in the packed data.

\subsubsection{Packing Data}

Packing data can be done in two ways: by using an intermediate typed buffer, or by serializing an object that already exists.

Consider a \var{Tree} ADT isomorphic to the one defined in Figure~\ref{fig:packed-definition}.
The code sample in Figure~\ref{fig:pack-with-write} uses the \var{Needs} intermediary buffer and functions like \var{startNode} and \var{writeLeaf} (generated automatically for the \var{Tree} type, see Section~\ref{sec:code-gen}).
Alternatively, the code snippet in Figure~\ref{fig:pack-with-pack} shows how to \textit{pack} a tree using the \var{pack} function. The buffers produced by these two approaches are identical.

\begin{figure}[ht]
	\begin{subfigure}[b]{0.49\textwidth}
\begin{lstlisting}[language=Haskell]
myPackedTree :: Packed '[Tree]
myPackedTree = finish needs 
    where needs = withEmptyNeeds $
        do
            startNode
            writeLeaf 1
            startNode
            writeLeaf 2
            writeLeaf 3
\end{lstlisting}
		\caption{Using the 'start' and 'write' functions}
		\label{fig:pack-with-write}
	\end{subfigure}
	\begin{subfigure}[b]{0.5\textwidth}
\begin{lstlisting}[language=Haskell, breaklines=true]
-- Defined in packed-data
pack :: Packable a => 
    a -> Packed '[a]
pack = ...

-- Generated automatically
instance Packable Tree where
    ...

myTree :: Tree
myTree = Node
	(Leaf 1)
	(Node (Leaf 2) (Leaf 3))

myPackedTree :: Packed '[Tree]
myPackedTree = pack myTree 
\end{lstlisting}
		\caption{Using the 'pack' function}
		\label{fig:pack-with-pack}
	\end{subfigure}
	\caption{How to pack a value with \packedhaskell}
\end{figure}

The \var{Needs} type represents intermediary buffers that \textit{needs} to be written to, before one can use them as packed data. Its definition in Figure~\ref{fig:needs-packed} is inspired by Bernardy et al.'s examples~\cite{linear-haskell}. \var{Needs} has two types parameters: the first one is the list (denoted \var{'[...]}) of the types of the values that the buffer expects before being considered \textit{ready}, and the second type argument is the types of the values packed in the buffer once reified. These parameters allow for type-safety when writing data and producing the final buffer. By using lists as type parameters, we can represent multiple values packed next to each other in a single buffer.

For example, \var{Needs '[Int] '[Tree]} represents a buffer that \textit{needs} a single integer, before it can be reified as a packed \var{Tree}. \var{Needs '[] '[Char, Int]} represents a buffer that is ready to be reified; it would contain a packed character, followed by a packed integer.

When a \var{Needs} is \textit{ready} or \textit{full} (i.e. its first type argument is an empty list \var{'[]}), we can produce the final packed buffer with \var{finish :: Needs '[] t -> Packed t}.

Finalised buffers take the form of a \var{Packed} value. This data type, defined in Figure~\ref{fig:needs-packed} has a single type parameter which tells the type of values packed inside.
For example, \var{Packed '[Tree, Tree]} is a buffer that contains two packed \var{Tree}s.

\begin{figure}[ht]
	\begin{lstlisting}[language=haskell]
newtype Needs p t = Needs ByteString

newtype Packed p = Packed ByteString 
	\end{lstlisting}
	\caption{Definition of the 'Needs' and 'Packed' data type}
	\label{fig:needs-packed}
\end{figure}

\subsubsection{Traversing Buffers}
\label{sec:traversing}

\begin{figure}[ht]
	\begin{subfigure}[b]{0.49\textwidth}
\begin{lstlisting}[language=Haskell, breaklines=true]
readInt :: PackedReader
    '[Int] r Int

sumInts :: PackedReader
    '[Int Int] '[] Int
sumInts = do
    i1 <- readInt
    i2 <- readInt
    return (i1 + i2)
\end{lstlisting}
		\caption{Reading Packed integers}
		\label{fig:packed-reader-example}
	\end{subfigure}
	\begin{subfigure}[b]{0.5\textwidth}
\begin{lstlisting}[language=Haskell, breaklines=true]
readChar :: PackedReader
    '[Char] r Char

example :: PackedReader
    '[Int, Int] '[] Int
example = do
    i1 <- readInt
    c <- readChar -- Type error
    return i1
\end{lstlisting}
		\caption{Invalid reading of packed values}
		\label{fig:packed-reader-example-error}
	\end{subfigure}\vspace{3mm}
	\begin{subfigure}[b]{1\textwidth}
\begin{lstlisting}[language=Haskell]
getSum :: Packed '[Int, Int] -> Int
getSum packed = runReader sumInts packed
\end{lstlisting}
		\caption{Execution of a PackedReader}
		\label{fig:packed-reader-exec}
	\end{subfigure}
	\caption{Reading packed values with \var{PackedReader}}
\end{figure}

We need a way to traverse packed data. To do so, we defined \var{PackedReader}, a computation modelling a cursor, \textit{reading} from left to right. \var{PackedReader} is parameterised by the type of the \var{Packed} data to read, the type of the data after the cursor has been shifted, and the type of the value produced by the computation. For example, in Figure~\ref{fig:packed-reader-example}, we define a function that computes the sum of two integers packed in the same buffer.

To illustrate the security that type-safety provides, consider the code snippet \ref{fig:packed-reader-example-error}, where we try to read a \var{Char} where the packed value is actually an integer. This code does not type-check, and an error will be raised at compile-time. This showcases how type-safety prevents programmers from reading values incorrectly and bad shifts of the \var{PackedReader}'s cursor.

\var{PackedReader} is actually an \textit{indexed} monad, where the type parameters represents the state of the computation \cite{indexed-monads}, i.e. the expected types of the data to traverse, and the types of the packed data after the cursor has been moved.

This allows us to leverage Haskell's type-system to ensure type-safe data manipulation, and the language's native syntax (e.g. \var{do}-notation) to give the code an imperative feel.
To actually execute a \var{PackedReader} computation, we use the \var{runReader} function. An example is given in Figure~\ref{fig:packed-reader-exec}.

The code generator described in Section~\ref{sec:code-gen} provides a 'case' function: a \var{PackedReader} which simulates pattern matching. Figure~\ref{fig:sum} is an example of a packed tree traversal, using the generated \var{caseTree} (defined in Figure~\ref{fig:case-tree-def}). This function computes the sum of the leaf values in a packed tree. The first lambda to pass (lines 4-7) is the function to execute if the packed tree is a \var{Leaf}, and the second one (lines 8-12) is for when the packed tree is a \var{Node}.

\begin{figure}[ht]
\begin{lstlisting}[language=Haskell, numbers=left]
sumTree :: PackedReader '[Tree] '[] Int
sumTree =
    caseTree
        ( do
            n <- reader
            return n
        )
        ( do
            left <- sumTree
            right <- sumTree
            return (left + right)
        )
\end{lstlisting}
	\caption{Computing the sum of the values in a Tree, using 'caseTree'}
	\label{fig:sum}
\end{figure}

\subsubsection{Code Generator}
\label{sec:code-gen}

Defining how to encode and decode packed data is a critical aspect of packed data manipulation. Since we deal with bytes, it is very easy to make mistakes, and tools like the type-system cannot be helpful here.
\packedhaskell takes away the programmer's responsibility to write encoding/decoding functions for their custom data types by providing a code generator.

Powered by Template Haskell \cite{DBLP:template-haskell}, this code generator defines the following for a given ADT:
\begin{itemize}
	\item An instance of the \var{Packable} typeclass, providing the \var{pack} function (seen in Figure~\ref{fig:pack-with-pack})
	\item An instance of the \var{Unpackable} typeclass, providing the \var{reader} function, to unpack a value from a packed stream.
	\item A 'case' function, to simulate pattern matching on a packed object (see Section~\ref{sec:traversing})
	\item A 'transform' function, similar to 'case'; it helps generate packed data out of \var{Packed} values (an example of its usage is given in Section~\ref{sec:example-increment-tree})
\end{itemize}

The code generator's entry point also takes as parameter a list of flags, which change its behaviour. Currently, these flags only deal with indirections (described in Section~\ref{sec:indirection}).

To invoke the code generator, the programmer has to call the \var{mkPacked} function. An example is given in Figure~\ref{fig:code-gen}.

\begin{figure}[ht]
\begin{lstlisting}[language=Haskell]
$(mkPacked ''Tree [InsertFieldSize])

\end{lstlisting}
	\caption{Invocation of the code generator}
	\label{fig:code-gen}
\end{figure}

\subsubsection{Indirections}
\label{sec:indirection}

The code generator accepts two flags:
\begin{itemize}
	\item \var{InsertFieldSize}: Inserts indirections before each field of the packed ADT.
	\item \var{SkipLastFieldSize}: Does not insert an indirection before the last field. We do not always need to jump over a final field, and not adding an indirection allows saving space in memory.
\end{itemize}

These flags will change the definition of the functions produced by the code generator. Figure~\ref{fig:indirections-api} illustrates these changes. Notice in Figure~\ref{fig:case-tree-def-ind} that each field is preceded by a \var{FieldSize} (a type alias to 32-bit integers we use as indirections), absent from the definition in Figure~\ref{fig:case-tree-def}, generated without passing indirection flags. This type transparency regarding indirections makes it easier to think about how to traverse a data structure: should we traverse everything to get to the nth field, or can we use indirections to jump there?

\begin{figure}[ht]
	\begin{subfigure}[b]{1\textwidth}
\begin{lstlisting}[language=Haskell, breaklines=true]
caseTree ::
    PackedReader '[FieldSize, Int] r a ->
    PackedReader '[FieldSize, Tree, FieldSize, Tree] r a ->
    PackedReader '[Tree] r a
\end{lstlisting}
		\caption{Definition of the 'case' function for Trees, when the 'InsertFieldSize' is on}
		\label{fig:case-tree-def-ind}
	\end{subfigure}\vspace{3mm}
	\begin{subfigure}[b]{1\textwidth}
\begin{lstlisting}[language=Haskell, breaklines=true]
caseTree ::
    PackedReader '[Int] r a ->
    PackedReader '[Tree, Tree] r a ->
    PackedReader '[Tree] r a
\end{lstlisting}
		\caption{Definition of the 'case' function for Trees, when no indirection flags are passed to the generator}
		\label{fig:case-tree-def}
	\end{subfigure}	\caption{Changes on the API caused by indirections}
	\label{fig:indirections-api}
\end{figure}

Consequently, code that manipulates packed data changes accordingly to the indirections flags given to the code generator. For example, consider Figure~\ref{fig:tree-traversals} where we traverse a binary tree to get its right-most value. In Figure~\ref{fig:tree-traversals-wo-ind}, we have to \textit{read} each left sub-trees to get to the right one. This is not optimal, as it causes a complete traversal of that sub-tree. However, in Figure~\ref{fig:tree-traversals-w-ind}, we \textit{skip} over a field using the \var{FieldSize} that precedes it. One drawback is that we have to explicitly skip \var{FieldSize}s we do not care about (before an integer value for example).

\begin{figure}[ht]
	\begin{subfigure}[b]{1\textwidth}
\begin{lstlisting}[language=Haskell, breaklines=true]
getRightMost :: PackedReader '[Tree] r Int
getRightMost =
    caseTree
        readInt
        ( do
            _ <- readTree
            getRightMost
        )
\end{lstlisting}
		\caption{Traversing a tree, without indirections}
		\label{fig:tree-traversals-wo-ind}
	\end{subfigure}\vspace{3mm}
	\begin{subfigure}[b]{1\textwidth}
\begin{lstlisting}[language=Haskell, breaklines=true]
getRightMost :: PackedReader '[Tree] r Int
getRightMost =
    caseTree
        ( do
            skipFieldSize -- Skipping leading indirection
            readInt
        )
        ( do
            skipWithFieldSize -- Skipping the left subtree
            skipFieldSize -- Skipping the right subtree's indirection
            getRightMost
        )
\end{lstlisting}
		\caption{Traversing a tree, with indirections}
		\label{fig:tree-traversals-w-ind}
	\end{subfigure}
	\caption{Tree traversals in \packedhaskell}
	\label{fig:tree-traversals}
\end{figure}

\subsection{Example: Incrementing values of a packed tree}
\label{sec:example-increment-tree}

The packed data produced by the library is immutable.
Since we are working with pointers and raw memory access under the hood, giving writing abilities (even through a monad) would break the purity and referential transparency of the library's API. Writing inside a buffer used by multiple \var{Packed} values would impact each of these variables, possibly leading to corruption.
Just like in functional languages, it is not possible to do in-place modifications on packed data; instead we have to duplicate it.

To do so, we can build a \var{Needs} value while traversing a packed value. This is what the generated 'transform' function is for: it behaves similarly to the 'case' function (in the sense that it helps with pattern matching on packed values, and it can be used recursively), but it forces its arguments to be \var{PackedReader}s that produce \var{Needs} values. It also helps the programmer by writing leading tags and inserting indirections when necessary.

Figure \ref{fig:increment-packed} gives an example of a function that takes a packed tree and produces another where all the leaf values are incremented by one.
We use the \var{transformTree} function (line 3) instead of \var{caseTree}, as the former invokes the \var{startLeaf}/\var{startNode} functions for us and thus helps keeping the code simple.
The first \var{PackedReader} (lines 4-7) passed to \var{transformTree} reads the integer of a packed \var{Leaf} and calls the \var{write} function from the \var{Packable} typeclass.
The second \var{PackedReader} (lines 8-12) calls the \var{incrementPacked} function recursively for each branch and builds back the packed \var{Node} using the \var{applyNeeds} function.

\begin{figure}[ht]
	\captionsetup{justification=centering}
\begin{lstlisting}[language=Haskell, numbers=left]
incrementPacked :: PackedReader '[Tree Int] r (Needs '[] '[Tree Int])
incrementPacked =
    transformTree
        ( do
            n <- readInt
            return (write (n + 1))
        )
        ( do
            left <- incrementPacked
            right <- incrementPacked
            return (applyNeeds left >> applyNeeds right)
        )
\end{lstlisting}
	\caption{Incrementing values of a packed tree}
	\label{fig:increment-packed}
\end{figure}

\subsection{Use case: Web Server}

To showcase the perks of using packed data, let's take the example of a very simple web API that has two endpoints '\var{/sum}' and '\var{/right-most}'. These endpoints respectively compute the sum of the leaves' values and get the right-most value in a packed tree passed to the request body.

Using the Scotty web framework\footnotemark{}\footnotetext{https://hackage.haskell.org/package/scotty}, we would implement the endpoints as in Figure~\ref{fig:packed-api-server}.
The endpoints are very simple: we get the packed tree, call the corresponding function \var{sumTree} and \var{getRightMost} (from Figures~\ref{fig:sum} and~\ref{fig:tree-traversals-wo-ind}) and send back the result.

The most interesting part of these endpoints is how we get the packed tree (on lines 4 and 9): Scotty's \var{body} function returns the raw request's body as a \textit{lazy} \var{ByteString}. However, a \var{Packed} value relies on a \textit{strict} \var{ByteString}. Thus, we need to use the \var{toStrict} function to do the conversion. This operation is $\mathcal{O}(n)$, but considering that a JSON-based API would traverse the entire \var{ByteString} to parse JSON as well, this is acceptable. \var{unsafeToPacked} is a $\mathcal{O}(1)$ operation to cast a \var{ByteString} into a typed \var{Packed} value.
Unlike a JSON-based API, we do not marshal the request body, saving both memory and time.

On the client side (Figure~\ref{fig:packed-api-client}), we enjoy a similar kind of optimisation, as the packed tree can be sent as-is (if the client already uses packed data). Additionally, both \var{fromPacked} and \var{fromStrict} (line 5) are $\mathcal{O}(1)$ operations, making the preparation of a request lightweight.

When running this server in parallel with one that uses \textit{unpacked} trees and JSON, we notice that the former runs noticeably faster than the later (Proper benchmarks on packed vs.\ unpacked data are done in Section~\ref{sec:bench}).

\begin{figure}[!ht]
	\begin{subfigure}[b]{\linewidth}
\begin{lstlisting}[language=Haskell, numbers=left]
endpoints :: ScottyM ()
endpoints = do
    get "/sum" $ do
        packedTree <- unsafeToPacked . toStrict <$> body
        sum <- liftIO $ runReader sumTree packedTree
        response sum

    get "/right-most" $ do
        packedTree <- unsafeToPacked . toStrict <$> body
        rightMost <- liftIO $ runReader getRightMost packedTree
        response rightMost
\end{lstlisting}
		\caption{Handling packed data in a web REST API Server}
		\label{fig:packed-api-server}
	\end{subfigure}
	\begin{subfigure}[b]{\linewidth}
\begin{lstlisting}[language=Haskell, numbers=left]
request methodGet "/sum" [] reqBody
  where
    packedTree :: Packed '[Tree]
    packedTree = ...
    reqBody = fromStrict (fromPacked packedTree)
\end{lstlisting}
		\caption{Sending packed data to a web REST API}
		\label{fig:packed-api-client}
	\end{subfigure}
	\caption{Packed data in a web REST API}
\end{figure}

\section{Evaluation}
\label{sec:eval}

To understand the impact of using the library to handle packed data, we evaluate it in two ways. First, we take a look at its usability and portability, by comparing how to manipulate packed data and unpacked data. Then, we look at the performance through a benchmark on simple tree traversals.

\subsection{Practical considerations}
\label{sec:eval-usability}

Since computations on packed data are lifted, it imposes limitations on how code can be written.
\packedhaskell tries to fill the gap between Haskell's features and what one might want to do with packed information, without modifying the compiler. However, this approach has some shortcomings.
Let's take a look at five aspects of our library-based approach to evaluate the usability of \packedhaskell.

\subsubsection{Type-safety and maintainability}
\label{sec:type-safety}

\packedhaskell provides a high-level and type-safe API for low-level data manipulation. It keeps the user from doing raw and unsafe pointer arithmetic by abstracting these operations behind a functional API.

As an example, consider Figures~\ref{fig:sum},~\ref{fig:haskell-raw-packed-sum} and~\ref{fig:c-packed-sum} where we compute the sum of the leaves of a packed tree using \packedhaskell and pointer arithmetic in Haskell and C.
In the later cases, dereferencing and pointer arithmetic bloats the code with multiple operations on numeric values that are not very expressive. In both languages, the type-checker will not be able to detect any pointer misuse.
On the other hand, when using \packedhaskell (Figure~\ref{fig:sum}), pointer arithmetic is hidden away behind functions like \var{reader}, \var{caseTree} and the \var{do} expressions, which guarantee correct pointer manipulation.
Therefore, our library makes code that manipulates packed data more readable, maintainable and memory-safe, compared to code that does raw pointer arithmetic.

\begin{figure}[!ht]
\begin{lstlisting}[language=Haskell]
sumPackedNonMonadic :: Ptr Word8  -> IO (Int, Word8)
sumPackedNonMonadic ptr = do
    tag <- peek ptr :: IO Word8
    let !nextPtr = ptr `plusPtr` 1
    case tag of
        0 -> do
            !n <- peek nextPtr
            let !nextPtr1 = plusPtr ptr 9
            return (n, nextPtr1)
        1 -> do
            (!left, !r) <- go (castPtr nextPtr)
            (!right, !r1) <- go r
            let !res = left + right
            return (res, r1)
        _ -> undefined



\end{lstlisting}
	\caption{Summing the values in a tree using raw pointer manipulation in Haskell}
	\label{fig:haskell-raw-packed-sum}
\end{figure}

\subsubsection{Portability}

\packedhaskell does not require modifying the compiler or the runtime system.
This means that the library's code does not deal with the compiler's implementation details. Thus, \packedhaskell can virtually be used with any modern version of GHC (as of today, the library has been tested with GHC versions 9.2 to 9.12). The library can be simply used as a third-party package, with no further constraints.

Not relying on the compiler's internal workings also means that \packedhaskell could be reimplemented in other languages with similar type systems like Scala, Rust or TypeScript.

\subsubsection{Code Generator's Applicability}

Thanks to its code generator (powered by meta-programming), \packedhaskell can be used with any user-defined algebraic data type, with the exception of unboxed types (denoted with a trailing \var{#}, like \var{ByteArray#}) and types that use them.

It also requires very little boilerplate code (see Figure~\ref{fig:code-gen} from Section~\ref{sec:code-gen}). Using a Template Haskell entry point like \var{mkPacked} is a common approach used by other popular Haskell libraries like Aeson\footnotemark{} for JSON encoding/decoding.\footnotetext{https://hackage.haskell.org/package/aeson}

\subsubsection{Generated Code and Flags}

We saw in Section~\ref{sec:indirection} that flags could change the definition of the functions created by the code generator, while the names of the generated functions do not change.
But what if we wanted to call the code generator twice for the same ADTs, passing different flags? This would lead to the definition of two functions with the same name (e.g.\ \var{caseTree}) in the same module, which is not allowed in Haskell. A workaround would be to call the code generator separately in dedicated modules, so that the different definitions are split, and we can import these modules according to our needs.

Another thing to consider is that the \var{Packed} data type does not us tell whether indirections are present in the packed data.
Therefore, if a user wants to build trees with and without indirections, it is up to them to use the correct 'case' and 'reader' functions generated with the correct flags.

We should also note that the same code cannot be reused with instances of the \var{Packable} typeclass generated with different indirection flags. Figures~\ref{fig:indirections-api} and~\ref{fig:tree-traversals} illustrate this problem: because of the strong typing of the packed data, \var{PackedReader}s need to explicitly skip over indirections (with \var{skipFieldSize}) when they are present.

\subsubsection{Pattern Matching}

One of Haskell's strengths is its powerful pattern matching support. Unfortunately, its syntax is not overload-able, meaning that we can not use it to pattern match on packed data.
As a solution, the code generator defines a 'case' function which takes lambdas as parameter, one for each constructor of the ADT (an example is given in Figure~\ref{fig:sum}).

However, an issue arises from not being able to use the standard syntax: we can't do nested pattern matching. In Haskell, we can pattern-match on a constructor's fields, while with our 'case' functions, pattern matching only considers constructors.
\\\\
\packedhaskell is a portable solution that provides type-safety for packed data manipulation. It is flexible (with flags for the code generator) and can be easily integrated to most modern Haskell projects. Even though the API sticks to Haskell's functional and pure paradigm, it does not support features like nested pattern matching.

\subsection{Performance}
\label{sec:bench}

We evaluate the performance of \packedhaskell by comparing the execution time of simple tree traversals with pointer-based Haskell, Gibbon, C and Golang.
We focus on four kinds of full tree traversals: computing the sum of values in a tree, evaluating an AST for a simple arithmetic language, getting the right-most node and incrementing the leaves' values. Table \ref{fig:bench} reports the execution times of these tree traversals.

We should note that we will not compare the size of the packed data with buffers created by Cap'n Proto. This is because \packedhaskell's serialisation format is very similar to Gibbon's, and Vollmer et al. have already demonstrated that the compiler's packing format is more space-efficient than Cap'n Proto \cite{local}.
For the same reason, we will not benchmark Cap'n Proto's traversals, as the same paper showcases that Gibbon's traversals are generally faster than Cap'n Proto. Thus, we can use Gibbon's performance to indirectly compare \packedhaskell with Cap'n Proto.

We should note that we will not compare the size of the packed data with buffers created by Cap'n Proto. This is because \packedhaskell's serialisation format is very similar to Gibbon's, and Vollmer et al. have already demonstrated that the compiler's packing format is more space-efficient than Cap'n Proto \cite{local}.
For the same reason, we will not benchmark Cap'n Proto's traversals, as the same paper showcases that Gibbon's traversals are generally faster than Cap'n Proto. Thus, we can use Gibbon's performance to indirectly compare \packedhaskell with Cap'n Proto.

For each benchmark, each tree is built symmetrically, meaning that a tree of 'size' 1 will have 2$^{1}$ = 2 leaves, a tree of 'size' 5 will have 2$^{5}$ = 32 leaves, etc.
The C and Haskell code were benchmarked using the Criterion library\footnotemark (with Haskell's Foreign Function Interface (FFI) to invoke the C code with negligible overhead)\footnotetext{https://hackage.haskell.org/package/criterion}. The Gibbon code was compiled with the \var{-p} and \var{--no-gc} options to use a packed layout and disable garbage collection. The Golang code was benchmarked using the \var{testing} library\footnotemark\footnotetext{https://pkg.go.dev/testing}. Code was compiled using GCC 11.4, GHC 9.10, Golang 1.18 and the latest Gibbon version as of March 2025. All benchmarks were run on an Intel machine with two Xeon Gold 6244 CPUs at 3.60 GHz, with 32Gb of RAM, running Ubuntu 22.04 LTS.

\begin{table}
	\small
	\caption{Execution times for summing the value in a tree}
	\label{fig:bench}
	\centering
	\captionsetup{justification=centerlast}
	\begin{subtable}[h]{\textwidth}\centering
	\begin{tabular}{|cc|r|r|r|r|r|}
		\hline
		\multicolumn{2}{|l|}{Language / Tree Size}         & \multicolumn{1}{c|}{1} & \multicolumn{1}{c|}{5} & \multicolumn{1}{c|}{10} & \multicolumn{1}{c|}{15} & \multicolumn{1}{c|}{20}           \\ \hline
		\multicolumn{2}{|c|}{C}                            & 7.56 ns                & 57.33 ns               & 1.76 us                 & 84.04 us                & 6.06 ms                           \\ \hline
		\multicolumn{2}{|c|}{Haskell}                      & 16.21 ns               & 202.68 ns              & 6.47 us                 & 207.17 us               & 6.56 ms                           \\ \hline
		\multicolumn{2}{|c|}{Gibbon}                       & 23.00 ns               & 359.00 ns              & 13.97 us                & 234.59 us               & 6.28 ms                           \\ \hline
		\multicolumn{1}{|c|}{\multirow{5}{*}{packed-data}} & \circled{1}            & 20.99 ns               & 150.31 ns               & 4.86 us                 & 157.94 us               & 5.14 ms \\ \cline{2-7}
		\multicolumn{1}{|c|}{}                             & \circled{2}            & 20.68 ns               & 152.82 ns               & 4.84 us                 & 156.19 us               & 5.22 ms \\ \cline{2-7}
		\multicolumn{1}{|c|}{}                             & \circled{3}            & 9.11 ns                & 133.68 ns               & 4.47 us                 & 143.39 us               & 4.68 ms \\ \cline{2-7}
		\multicolumn{1}{|c|}{}                             & \circled{4}            & 8.95 ns                & 134.16 ns               & 4.54 us                 & 145.97 us               & 4.91 ms \\ \hline
		\multicolumn{2}{|c|}{Golang}                       & 5.30 ns                & 111.30 ns              & 3.86 us                 & 165.05 us               & 16.49 ms                          \\ \hline
	\end{tabular}
	\caption*{
		\circled{1}: Without indirections.
		\circled{2}: With indirections.
		\\
		\circled{3}: Non-monadic, without indirections.
		\circled{4}: Non-monadic, with indirections.
	}

	\caption{Execution times for summing the value in a tree}

\end{subtable}

	\begin{subtable}[h]{\textwidth}\centering
	\begin{tabular}{|cc|r|r|r|r|r|}
		\hline
		\multicolumn{2}{|l|}{Language / Tree Size}         & \multicolumn{1}{c|}{1} & \multicolumn{1}{c|}{5} & \multicolumn{1}{c|}{10} & \multicolumn{1}{c|}{15} & \multicolumn{1}{c|}{20}             \\ \hline
		\multicolumn{2}{|c|}{C}                            & 10.74 ns               & 121.19 ns              & 4.65 us                 & 181.40 us               & 9.24 ms                             \\ \hline
		\multicolumn{2}{|c|}{Haskell}                      & 16.22 ns               & 278.23 ns              & 8.67 us                 & 280.97 us               & 13.20 ms                            \\ \hline
		\multicolumn{2}{|c|}{Gibbon}                       & 73.00 ns               & 723.00 ns              & 10.60 us                & 181.50 us               & 5.34 ms                             \\ \hline
		\multicolumn{1}{|c|}{\multirow{3}{*}{packed-data}} & \circled{1}            & 24.08 ns               & 168.80 ns               & 5.25 us                 & 168.18 us               & 5.41 ms   \\ \cline{2-7}
		\multicolumn{1}{|c|}{}                             & \circled{2}            & 27.84 ns               & 471.86 ns               & 15.69 us                & 751.96 us               & 111.96 ms \\ \cline{2-7}
		\multicolumn{1}{|c|}{}                             & \circled{3}            & 10.89 ns               & 151.21 ns               & 5.11 us                 & 164.25 us               & 5.27 ms   \\ \hline
		\multicolumn{2}{|c|}{Golang}                       & 5.29 ns                & 108.90 ns              & 4.42 us                 & 180.51 us               & 15.41 ms                            \\ \hline
	\end{tabular}
	\caption*{
		\circled{1}: Without indirections.
		\circled{2}: Unpacking, then traverse.
		\circled{3}: Non-monadic, without indirections.
	}
	\caption{Execution times for evaluating a packed AST for arithmetic expressions}
\end{subtable}

	\begin{subtable}[h]{\textwidth}\centering
	\begin{tabular}{|cc|r|r|r|r|r|}
		\hline
		\multicolumn{2}{|l|}{Language / Tree Size}         & \multicolumn{1}{c|}{1} & \multicolumn{1}{c|}{5} & \multicolumn{1}{c|}{10} & \multicolumn{1}{c|}{15} & \multicolumn{1}{c|}{20}             \\ \hline
		\multicolumn{2}{|c|}{C}                            & 7.35 ns                & 9.20 ns                & 10.75 ns                & 28.23 ns                & 86.15 ns                            \\ \hline
		\multicolumn{2}{|c|}{Haskell}                      & 11.63 ns               & 15.52 ns               & 19.45 ns                & 23.33 ns                & 27.30 ns                            \\ \hline
		\multicolumn{2}{|c|}{Gibbon}                       & 29.00 ns               & 39.00 ns               & 62.00 ns                & 93.00 ns                & 74.00 ns                            \\ \hline
		\multicolumn{1}{|c|}{\multirow{5}{*}{packed-data}} & \circled{1}            & 21.68 ns               & 93.63 ns                & 2.80 us                 & 95.27 us                & 3.24 ms   \\ \cline{2-7}
		\multicolumn{1}{|c|}{}                             & \circled{2}            & 21.67 ns               & 25.85 ns                & 32.36 ns                & 43.06 ns                & 112.77 ns \\ \cline{2-7}
		\multicolumn{1}{|c|}{}                             & \circled{3}            & 20.94 ns               & 215.35 ns               & 6.60 us                 & 349.04 us               & 67.94 ms  \\ \cline{2-7}
		\multicolumn{1}{|c|}{}                             & \circled{4}            & 9.29 ns                & 77.98 ns                & 2.82 us                 & 95.95 us                & 3.32 ms   \\ \cline{2-7}
		\multicolumn{1}{|c|}{}                             & \circled{5}            & 8.12 ns                & 10.54 ns                & 16.52 ns                & 26.86 ns                & 94.32 ns  \\ \hline
		\multicolumn{2}{|c|}{Golang}                       & 3.45 ns                & 8.58 ns                & 18.25 ns                & 26.61 ns                & 35.05 ns                            \\ \hline
	\end{tabular}

	\caption*{
		\circled{1}: Without indirections.
		\circled{2}: With indirections.
		\circled{3}: Unpacking, then traverse.
		\\
		\circled{4}: Non-monadic, without indirections.
		\circled{5}: Non-monadic, with indirections.
	}
	\caption{Execution times for getting the right-most value in a tree}
\end{subtable}

	\begin{subtable}[h]{\textwidth}\centering
	\begin{tabular}{|cc|r|r|r|r|r|}
		\hline
		\multicolumn{2}{|l|}{Language / Tree Size}         & \multicolumn{1}{c|}{1} & \multicolumn{1}{c|}{5} & \multicolumn{1}{c|}{10} & \multicolumn{1}{c|}{15} & \multicolumn{1}{c|}{20}             \\ \hline
		\multicolumn{2}{|c|}{C}                            & 7.33 ns                & 35.45 ns               & 1.79 us                 & 82.26 us                & 7.68 ms                             \\ \hline
		\multicolumn{2}{|c|}{Haskell}                      & 19.50 ns               & 294.18 ns              & 10.55 us                & 501.31 us               & 75.83 ms                            \\ \hline
		\multicolumn{2}{|c|}{Gibbon}                       & 71.00 ns               & 635.00 ns              & 30.40 us                & 486.78 us               & 14.42 ms                            \\ \hline
		\multicolumn{1}{|c|}{\multirow{3}{*}{packed-data}} & \circled{1}            & 41.99 ns               & 469.68 ns               & 16.80 us                & 648.15 us               & 74.50 ms  \\ \cline{2-7}
		\multicolumn{1}{|c|}{}                             & \circled{2}            & 60.29 ns               & 821.67 ns               & 33.45 us                & 1.99 ms                 & 245.64 ms \\ \cline{2-7}
		\multicolumn{1}{|c|}{}                             & \circled{3}            & 42.57 ns               & 580.03 ns               & 21.97 us                & 1.03 ms                 & 165.04 ms \\ \hline
		\multicolumn{2}{|c|}{Golang}                       & 6.61 ns                & 129.60 ns              & 4.29 us                 & 179.44 us               & 16.97 ms                            \\ \hline
	\end{tabular}
	\caption*{
		\circled{1}: Using NeedsBuilder.
		\circled{2}: Unpacking, increment and repack.
		\\
		\circled{3}: Deserialise and increment, and repack.}
	\caption{Execution times for incrementing the values in a tree}
\end{subtable}

	\captionsetup{}
\end{table}

In this section, we will ignore the benchmark cases prefixed with \textit{Non-monadic}, we will get back to them in Section~\ref{sec:reflection}.

For summing the values in a tree, using \packedhaskell can provide up to a 20\% speed-up compared to native Haskell. It also beats C by running 15\% faster. We can note that, for this specific scenario, having indirections slightly slows down the traversal (around 2\%): we do not need them to optimise the traversal, so we skip over them, leading to additional smaller jumps.

Evaluating an AST is similar to summing the values of a tree, as it requires a full traversal of a tree, without skipping any fields. The AST used for this benchmark is one for a small arithmetic language with integer values, addition, subtraction, multiplication, and division. This means that we will not have to deal with two constructors/tags like for the previous benchmark (for leaves and nodes), but five (one for the value, and one for each arithmetic operation).
Similarly to the previous benchmark, using \packedhaskell provides a 60\% speed-up compared to regular Haskell.
The \textit{Unpacking, then traverse} line shows the runtime for functions that unpacks the packed AST, and then evaluates it. This case shows that doing so leads to an 8.5x slowdown compared to native Haskell and illustrates the cost of unpacking a value before processing it, i.e. that \var{time(unpacking) + time(eval_unpacked) > time (eval_packed)}.

The third benchmark focuses on getting the right-most value in a tree. Unlike the previous cases, here a lot of jumps will be done, as we do not need to go through each branch of the tree. However, using \packedhaskell, the operation is 10000 times slower than regular Haskell, as we need to traverse the entire packed tree to reach the last value. But this benchmarks illustrates the benefits of indirections: when they are present, we can skip over subtrees, avoiding a lot of useless traversals. Indirections allow this computation to be only four times slower than unpacked Haskell.

In the final benchmark, we evaluate the performance of incrementing the leaves' values in a tree. Note that the C and Golang benchmarks increment the trees \textit{in-place}, while the others create a new tree.
Here, Haskell is much slower than C, since Haskell's functional paradigm forces such operations to produce an entirely new tree. In \packedhaskell, this can be done by using the \var{traverse} function and calling the \var{finish} function. The performance of \packedhaskell are surprisingly comparable with regular Haskell. This can be justified by the fact that \packedhaskell only does one big memory allocation for the entire buffer, instead of one per tree node.
In the \textit{Unpacking, increment and repack} case, we deserialise the packed tree, increment the leave's values (using the same function as for the native Haskell case), and then repack the tree.
The \textit{Deserialise and increment, and repack} case produces an unpacked tree already incremented (i.e. the leave's values are incremented in the 'case' function's lambdas) and then re-serialiases the tree. The latter case has better performance than the former, as it avoids an additional traversal and the creation of a new unpacked tree.
Again, these last two cases show that it is faster to use unpacked data as-is, instead of deserialising it first, whether it is for read-only or writing operations.

As a final note, we observed via some preliminary benchmarking that the packed data approach appears to perform worse on Apple ARM devices, where the non-packed (pointer-chasing) code is generally faster. We speculate that this may be due to ARM either handling pointer-chasing within a tree better, or punishing unaligned access more severely. Full investigation of packed data performance on ARM architecture is future work. Regardless, even on ARM, there was a significant benefit to operating directly on packed data rather than the "unpack, process and repack" round-trip.

We conclude that \packedhaskell can provide better performance than native Haskell, but not in every case. In the best scenarios, the speed-ups can reach 60\%, compared to unpacked data, but in the worst scenarios, it is considerably slower (from 4 to 10000 times). In the next section, we speculate on why sometimes the library performs slower than native Haskell.

\section{Reflection}
\label{sec:reflection}

\begin{figure}[!ht]
	\begin{subfigure}[b]{\textwidth}
\begin{lstlisting}[language=Haskell]
evalPacked :: PackedReader '[AST] r Int32
evalPacked =
    caseAST
        reader
        (opLambda (+))
        (opLambda (-))
        (opLambda (*))
        (opLambda div)
  where
    {-# INLINE opLambda #-}
    opLambda ::
        (Int32 -> Int32 -> Int32) ->
        PackedReader '[AST, AST] r Int32
    opLambda f = R.do
        left <- evalPacked
        right <- evalPacked
        R.return (f left right)
\end{lstlisting}
		\caption{Using \packedhaskell}
		\label{fig:ast}
	\end{subfigure}
	\begin{subfigure}[b]{\textwidth}
\begin{lstlisting}[language=Haskell]
evalPackedNonMonadic :: Packed (AST ': r) -> IO Int32
evalPackedNonMonadic packed = fst <$> go (unsafeForeignPtrToPtr fptr)
  where
    (BS fptr _) = fromPacked packed
    go :: Ptr Word8 -> IO (Int32, Ptr Word8)
    go ptr = do
        tag <- peek ptr :: IO Word8
        let !nextPtr = ptr `plusPtr` 1
        case tag of
            0 -> do
                !n <- peek nextPtr :: IO Int32
                return (fromIntegral n, plusPtr nextPtr (sizeOf n))
            1 -> opLambda (+) nextPtr
            2 -> opLambda (-) nextPtr
            3 -> opLambda (*) nextPtr
            4 -> opLambda div nextPtr
            _ -> undefined
    {-# INLINE opLambda #-}
    opLambda :: (Int32 -> Int32 -> Int32) -> Ptr Int32 -> IO (Int32, Ptr Word8)
    opLambda f ptr = do
        (!left, !r) <- go $ castPtr ptr
        (!right, !r1) <- go r
        let !res = left `f` right
        return (res, r1)
\end{lstlisting}
		\caption{Using pointer arithmetic}
		\label{fig:ast-raw}
	\end{subfigure}
	\caption{Evaluating a packed AST}
	\label{fig:monadic-vs-raw}
\end{figure}

We concluded in the previous section that the performance of \packedhaskell were satisfactory. This can be explained by the fact that we make the most of the L1 cache and do not jump around using pointers. In this section we consider an alternative way of writing code in Haskell for packed data that can provide better performance than \packedhaskell. This approach, however, is far less maintainable and safe.

In Haskell, monads are not defined at the compiler level, they are regular data structures defined in Haskell's \var{base} standard library. Therefore, even though they are a big part of the functional paradigm, they do not enjoy any dedicated compiler optimisation.

To manipulate packed data, \packedhaskell uses two monads: \var{IO}, required to do any kind of low-level operation (like dereferencing a pointer using \var{peek}), and \var{PackedReader}, which provides the high-level API for packed data traversals.

Since monadic operations can be as optimised as any other kind of computation (e.g. through inlining), inevitably, using two monads on top of each other like \packedhaskell must lead to some computing overhead.

We argued in Section~\ref{sec:type-safety} that \packedhaskell's API allowed for readable and type-safe code. Figure \ref{fig:ast-raw} is an example of what traversing packed data without \packedhaskell would look like. It is safe to say that this code is way less pleasant to read and harder to maintain than with the library (Figure~\ref{fig:ast}). However, it does get rid of one level of computation induced by the \var{PackedReader} monad.

Benchmarks using this non-monadic approach are included in Table~\ref{fig:bench} and labelled as \textit{Non-monadic}.
In the summing and AST evaluation benchmarks, this approach yields slightly better performance (between 2 and 6\% compared to the monadic approach).
However, in the third benchmark, with indirections, it is 19\% faster than its \packedhaskell counterpart. But it is still 3.5 times slower than regular Haskell, possibly because this computation relies heavily on following indirections and jumping around in the serialized tree. This leads to an extensive use of the \var{IO} monad, slowing down the traversal.

This confirms that our monadic approach to packed data manipulation causes some kind of computing overhead. However, this is specific to Haskell. Implementations for other languages might not suffer from this, but they should be able to leverage similar performance enhancements. As future work, we could use staging and compile-time evaluation to eliminate that computing overhead.

\section{Future Work}
\label{sec:future-work}

While \packedhaskell is a working library, it can be improved in many ways; or it can serve as a stepping stone for further research on the topic. In this section, we enumerate possible future work that can be done on the library or in this research domain.

Building packed data (through a \var{Needs} buffer) is slow (four times slower than building native objects).
Internally, it relies on the \var{bytestring-strict-builder} library\footnotemark\footnotetext{https://hackage.haskell.org/package/bytestring-strict-builder} which, in itself provides good performance\footnotemark{}\footnotetext{https://github.com/haskell-perf/strict-bytestring-builders}. However, as for packed data traversals, building packed data is implemented through a monad, \var{NeedsBuilder}. As discussed in Section~\ref{sec:reflection}, even though it provides a functional and type-safe API, such approach can cause some computing overhead.
A future version of \packedhaskell could focus on making the packing process faster.

\packedhaskell only supports its own packing layout. However, some systems could take advantage of data packed with a custom or pre-defined layout. Take JSON for example: the client of an API that serves JSON data would not have to deserialise the server's response, and could use it as is. This is also applicable for web browsers and HTML, especially since web pages can be sizeable.
A next step for \packedhaskell would be to handle data packed with a more complex layout like JSON, HTML or CBOR\footnotemark \footnotetext{https://cbor.io}.

We have seen that accessing a field in a packed structure is not a straightforward task: we have to either use indirections or traverse the preceding fields. We could draw inspiration from Chilimbi, Davison \& Larus' work~\cite{DBLP:conf/pldi/ChilimbiDL99} and reorder fields according to their access pattern, moving the most accessed ones first in a packed structure \cite{marmoset}. This would help programs access 'popular' fields more easily, with fewer jumps in the packed buffer.

In \packedhaskell, indirections take the form of a 32-bit integer. We could definitely save some space in the packed buffer by choosing the size of this indirection accordingly to the following field. For example, we do not need a 32-bit indirection for a four-characters string, one byte would be enough.
To go even further, some packed data do not need indirections at all, if their size can be guesses statically (e.g.\ a packed character will always be one byte, a 16-bit number two bytes, etc.). In that case, indirections are useless. We can avoid inserting them (reducing the size of the packed buffer) and deduce the size of that packed data by relying on the type-checker and a couple of typeclasses.

As mentioned in Section~\ref{sec:eval-usability}, we could also allow the same code on packed data to be reused whether the data has been packed with or without indirections.

Another area of improvement would be not to insert indirections in the packed buffer itself, but in a separate heap object. This would allow for packed data to be smaller and enable parallelism in tree traversals.

Finally, because of the data layout used by \packedhaskell, memory access can be unaligned. This noticeably affects performance on ARM CPUs. In the future, \packedhaskell could include a flag that forces the insertion of padding in the packed buffer, to align tags, indirections and fields.
\\\\
The portability of \packedhaskell across other languages remains theoretical. While its implementation does not rely on modifying the compiler or the runtime system, it does heavily rely on Haskell's type-system. It would be interesting to implement \packedhaskell in an other typed language like Scala, TypeScript or Rust.
\\\\
We concluded in Section~\ref{sec:reflection} that our library-based approach in Haskell suffered from a computing overhead caused by the \var{IO} monad, blocking any possible future enhancement of the library's performance.
Deeply-embedded domain-specific languages (EDSLs) could be a solution to this problem. EDSLs expose functions that do not actually execute an operation, but instead produce an AST \cite{deep-vs-shallow-edsl}. These ASTs can then be optimised \cite{meta-repa} and compiled into a full-blown program in the host language or produce a standalone executable \cite{feldspar}. A few EDSLs have performance as their main objective \cite{meta-repa}, and to reach it, it's not uncommon to rely on a backend written in a lower-level language, like C \cite{copilot}.

To provide better performance, \packedhaskell could be reimplemented as a deeply embedded DSL: at compile time, using Template Haskell, we could produce an AST and generate C code. Using Template Haskell's quotation feature, we could inject the call to that C code into the Haskell program, using the language's C FFI. This way, we would still have a type-safe and functional API, while enjoying C's fast performance \cite{compiling-embedded-languages}. Furthermore, this C code could be reused as a common backend for implementations of \packedhaskell in other languages.

\section{Conclusion}

Exchanging data using a packed format can be faster than when using a human-readable format like JSON, as the recipients can use the serialised data as-is, avoiding any deserialisation step. Manipulating packed data can also help save computation time, as it avoids chasing pointers. However, this format is not easy to adopt, as most implementations require either compiler modifications or a dedicated compiler.

In this paper, we investigate library-based support for packed data through \packedhaskell. This Haskell library provides a type-safe and functional API that allows packing data, processing it, and unpacking it. It relies on Template Haskell and Haskell's type-system, meaning that this library could virtually be implemented in any language that supports meta-compilation and higher-kinded types.

Our benchmarks show that the library can leverage better performance than native Haskell (up to 60\%), but in some scenarios like getting the right-most value in a tree, \packedhaskell (with indirections) is four times slower. This is due to a computation overhead caused by the monadic approach used in the implementation, as benchmarks show that 'pure' pointer arithmetic-based implementations can provide more competitive performance.

While a library-based approach for packed data support is feasible and portable, there is a performance and usability tradeoff to consider.
We think that \packedhaskell would benefit from staging and compile-time optimisations, for example, through an EDSL, which would generate and call C code. This would help get rid of computing overhead caused by the host language, and could be reused to build a common backend used by implementations of similar libraries in other languages.

\bibliography{packed}

\end{document}